\documentclass[11pt a4paper]{article}
\usepackage{jheppub}

\usepackage{amssymb}
\usepackage{graphicx}
\usepackage{subfigure}
\usepackage{url}
\usepackage{epstopdf}

\title{Estimation of Inflation parameters for Perturbed Power Law model using recent CMB measurements}

\author [\dagger]{Suvodip Mukherjee,}
\author [\dagger]{Santanu Das,}
\author [\ddagger]{Minu Joy}
\author [\dagger]{and Tarun Souradeep}
\affiliation[\dagger]{Inter University Centre for Astronomy and Astrophysics, \\
Post Bag 4, Ganeshkhind, Pune-411007, India}
\affiliation[\ddagger]{Dept. of Physics, Alphonsa College,
Pala 686574, India}

\emailAdd{suvodip@iucaa.ernet.in}
\emailAdd{santanud@iucaa.ernet.in}
\emailAdd{minujoy@gmail.com}
\emailAdd{tarun@iucaa.ernet.in}

\abstract{Cosmic Microwave Background (CMB) is an important probe for understanding the inflationary era of the Universe. We consider the Perturbed Power Law (PPL)  model of inflation which is a soft deviation from Power Law (PL) inflationary model. This model captures the effect of higher order derivative of Hubble parameter during inflation, which in turn leads to a non-zero effective mass $m_{\rm eff}$ for the inflaton field. The higher order derivatives of Hubble parameter at leading order sources constant difference in the spectral index for scalar and tensor perturbation going beyond PL model of inflation. PPL model have two observable independent parameters, namely spectral index for tensor perturbation $\nu_t$ and change in spectral index for scalar perturbation $\nu_{st}$ to explain the observed features in the scalar and tensor power spectrum of perturbation. From the recent measurements of CMB power spectra by WMAP, Planck and BICEP-2 for temperature and polarization, we estimate the feasibility of PPL model with standard $\Lambda$CDM model. Although BICEP-2 claimed a detection of $r=0.2$, estimates of dust contamination provided by Planck have left open the possibility that only upper bound on $r$ will be expected in a joint analysis. As a result we consider different upper bounds on the value of $r$ and show that PPL model can explain a  lower value of tensor to scalar ratio ($r<0.1$ or $r<0.01$) for a scalar spectral  index of $n_s=0.96$ by having a non-zero value of effective mass of the inflaton field $\frac{m^2_{\rm eff}}{H^2}$. The analysis with WP+ Planck likelihood shows a non-zero detection of $\frac{m^2_{\rm eff}}{H^2}$ with $5.7\,\sigma$ and $8.1\,\sigma$ respectively for $r<0.1$ and $r<0.01$. Whereas, with BICEP-2 likelihood $\frac{m^2_{\rm eff}}{H^2} = -0.0237 \pm 0.0135$ which is consistent with zero.} 

\begin{document}
\maketitle

 \section{Introduction}
The precision measurement of temperature and polarization of Cosmic Microwave Background (CMB) by several experiments like WMAP, Planck, BICEP etc. have enabled us to constraint several cosmological parameters with unprecedented accuracy. Minimal $\Lambda$CDM model explains the observed temperature spectra by Planck \cite{Planck_1, Planck_param}. The recent detection of $B$ mode polarization of CMB by BICEP \cite{BICEP_1} have provided us a window for  measuring the primordial gravitational waves which is an important probe to understand the nature of inflation. 

Inflation is the rapid accelerated expansion of the Universe postulated at very early time and that also predicts the generation of initial scalar and tensor perturbation from the quantum fluctuations of the early universe. These perturbations lead to anisotropies in CMB temperature field and also provide the initial seed for structure formation of the Universe. CMB power spectra for temperature and polarization is a window to the early era of the Universe and dynamics of the inflationary era can be studied with several observable quantities from CMB power spectra. Many single field inflationary models predict adiabatic, Gaussian and nearly scale independent perturbation and recent measurements from WMAP \cite{wmap1,wmap2} and Planck \cite{planck_infla} are consistent with these prediction. A detailed study for several single field inflationary model with the recent data from WMAP, Planck and BICEP-2 are done by Martin et al. \cite{martin_1, martin_2, martin_3}. One of such model is Power Law (PL) inflation introduced by Lucchin \& Matarrese \cite{lucchin} 
where scale factor $a(t)$ during inflation evolves as $a(t) \sim t^p$. It predicts a scale invariant power spectrum for both scalar and tensor perturbation. The spectral indices are related by $n_s-1=n_t \propto (\frac{d\ln \rm H}{d\phi})^2$. The consistency relation between tensor to scalar ratio $r$ and tensor spectral index $n_t$ is $n_t= -r/6.2$ \cite{lyth}.  However, for the inflation to end and reheating to begin, it is essential to consider the change in spectral index by $n_{run}= dn_s/d\ln k$. Souradeep et al. \cite{ts} extended the PL model to a model called  Perturbed Power Law (PPL), which considers soft deviation from PL inflationary model by capturing the next higher order derivatives of Hubble parameter. 

In this paper, we briefly discuss the PPL model with two main  parameters $\nu_t$ and $\nu_{st}$ to capture the effect of $n_s$, $n_t$ and $n_{run}$. These two parameters are related to the inflationary parameters $(\frac{d\ln \rm H}{d\phi})^2$ and $\frac{d^2\ln \rm H}{d\phi^2}$. The model also predicts  similar consistency relation between $r$ and $n_t$. Considering the power spectra measured by WMAP and Planck and also the $B$ mode polarization recently measured by BICEP-2 \cite{BICEP_1}, we obtain the constraints on these inflationary parameters.

The paper is organized as follows. In Sec. \ref{ppl}, we briefly discuss the basic formalism of PL and PPL models. In Sec. \ref{cp}, we constrain the inflationary parameters $\nu_s, \, \nu_t$ and $\nu_{st}$ using the WMAP, Planck and  BICEP-2 likelihood. Conclusion and feasibility of PPL model is discussed in Sec. \ref{conc}.

\section{Background of Perturbed Power Law (PPL) model of inflation}\label{ppl}
Perturbed Power Law (PPL) \cite{ts, minu} is an extension to Power Law (PL) inflationary model \cite{lucchin, vs_3, vs_2, ts_thesis} by considering higher order corrections to Hubble parameter. These corrections to Hubble parameter leads to an effective mass for the inflaton field, which in turn affects the density power spectra $P(\vec k)$.

For single field inflationary models, Hubble parameter $H(k)$ and its evolution during inflation can be translated into the inflationary potential using Hamilton-Jacobi formulation \cite{salopek, lidsey} as,

\begin{eqnarray}\label{hj}
V(\phi)= \frac{3m_p^2\rm H^2(\phi)}{8\pi}\big[ 1- \frac{m_p^2}{12\pi}\big(\frac{\partial \ln \rm H}{\partial \phi}^2\big)\big],
\end{eqnarray}  
where, $m_p$ is the Planck mass. 

For any given potential, slow-roll parameters ($\epsilon$ and $\delta$) are related to the derivatives of the Hubble parameter as
\begin{eqnarray}\label{slow}
\epsilon=& -\frac{\dot {\rm H}}{\rm H^2} = \frac{m_p^2}{4\pi}(\frac{d\ln \rm H}{d\phi})^2\,\,<<1,\\
\delta=& \frac{\ddot \phi}{\dot \phi H}= -\frac{m_p^2}{4\pi \rm H}\frac{d^2\rm H}{d\phi^2}\,\,<<1.
\end{eqnarray}
These parameters are assumed to be less than 1 (slow-roll approximation, $\ddot \phi << 3H\dot \phi$) and are also validated by Planck \cite{Planck_param}. These parameters are used to study the dynamics of inflaton field and are related to several observable quantities like scalar spectral index ($n_s$) and tensor spectral index ($n_t$) and also higher order corrections to these spectral index.

For PL model, the potential takes the form,
\begin{equation}\label{poten}
V(\phi)= V_0 \exp(-\sqrt{\frac{4\pi}{p}}\frac{\phi}{m_p}).
\end{equation}
This leads to slow-roll parameters $\epsilon = -\delta = 1/p$ and corresponds to spectral index for scalar perturbation $(\nu_s)$ and tensor perturbation $(\nu_t)$ as
\begin{equation}\label{nsst}
\nu_s = \nu_t = \frac{3}{2} -\frac{\nu}{2},
\end{equation}
where, 
\begin{equation}\label{nust1}
\nu= \frac{-2 \epsilon}{1-\epsilon}.
\end{equation}
The parameter $\nu$ is related to the usual definition of spectral indices $n_s$ and $n_t$ by
\begin{equation}\label{nsnt}
n_s -1= n_t = \nu.
\end{equation}
This constant spectral index for both scalar and tensor power spectra suggest its name as 'Power Law' (PL) model of inflation. Tensor to scalar ratio $r$ is related to the spectral index by \cite{lyth},
\begin{equation}\label{r_nt}
r=-6.2n_t.
\end{equation}
For a single field inflation model, effective mass $m_{\rm eff}$ of the inflaton field can be defined as
\begin{equation}\label{nsme}
\frac{m_{\rm eff}^2}{\rm H^2} = -(\nu_{st})(\delta+3) + \frac{\dot \epsilon - \dot \delta}{\rm H},
\end{equation}
where, $\nu_{st}= (\epsilon +\delta)$. It can be expressed by higher order derivative of Hubble parameter and the leading order contribution is from the $2^{nd}$ derivative of Hubble parameter, expressed as
\begin{equation}\label{meff}
\frac{m_{\rm eff}^2}{\rm H^2} \approx \frac{3m_p^2}{4\pi} \frac{d^2\ln \rm H}{d\phi^2}. 
\end{equation}

For the PL model of inflation, $m_{\rm eff}= 0$ and hence leads to constant spectral index for scalar and tensor perturbation as $n_s-1=n_t \approx -2\epsilon$. A non-zero effective mass $m_{\rm eff}$ affects only the scalar perturbation, whereas tensor perturbation continues to be massless excitations. 

We consider soft departure from the PL model  by accounting for non-zero value of $m_{\rm eff}^2$, which 
leads to change in the slow roll parameters and varying spectral index for scalar perturbation. For PPL model of inflation, spectral index for scalar perturbation $\nu_s$ and tensor perturbation $\nu_t$ are defined as
\begin{align}\label{spec}
\nu_{s} &= \frac{3}{2} + \epsilon(k) + \nu_{st},\\
\nu_{t} &= \frac{3}{2} + \epsilon,
\end{align}
where, $\nu_{st}\approx \nu_s - \nu_t$ is the difference between spectral index for scalar and tensor perturbation arising due to non-zero value of $\frac{d^2\ln \rm H}{d\phi^2}$. So in the model of PPL we can express the spectral features of primordial power spectrum for both scalar and tensor by two parameters $\nu_t$ and $\nu_{st}$.

In the presence of effective mass, the evolution of slow roll parameter $\epsilon$ can be expressed as
\begin{equation}\label{der_eps}
\dot \epsilon = 2\epsilon\nu_{st}\rm H.
\end{equation}
By solving this equation upto leading order in $\epsilon$, we find $\epsilon(k)$ as
\begin{equation}\label{nust_eps}
\epsilon (k) \propto k^{2\nu_{st}},
\end{equation}
where $\nu_{st}$ is a constant. 
The power spectrum for scalar perturbation $P_s(k)$ and tensor perturbation $P_t(k)$ becomes,
\begin{align}\label{pk_1}
P_s(k) &= A(\nu_{t},\nu_{s})\bigg(\frac{\rm H^2(k)}{\epsilon(k)}\bigg),\\ 
P_t(k) &= 8A(\nu_{t},\nu_{t})\rm H^2(k). 
\end{align}
These above equations can be written in terms of the wave number $k$ as, 
\begin{align}\label{pk}
P_s(k) = A(\nu_{t},\nu_{s})k^{3- 2\nu_{s}},\\ 
P_t(k) = 8A(\nu_{t},\nu_{t})k^{3- 2\nu_{t}}, 
\end{align}
where, $A(x,y) = \frac{4^y \Gamma^2(y)}{8\pi^3(x-1/2)^{2y-1}}$. 
In PPL, we can understand the evolution of Hubble parameter from $\nu_t$ and $\nu_{st}$ which are related to the $1\rm st$ and $2\rm nd$ order derivative of Hubble parameter 
and can completely capture the essence of  $n_s$ and $n_t$ and $n_{run}= dn_s/d\ln k$. 

\section{Estimation of Cosmological parameters for Perturbed Power Law}\label{cp}
PPL model discussed in the previous section shows that $\nu_t$ and $\nu_{st}$ are the observables to study the evolution of Hubble parameter during inflation. 
Using the results of CMB temperature and polarization field from several  missions like WMAP, Planck and BICEP, we want to estimate these parameters. However, 
recently Planck has made an estimation of B mode polarization due to dust foregrounds at $150 \,\rm GHz$ in the BICEP patch of the sky and have shown that the dust foreground contamination considered by BICEP \cite{planck_dust} is underestimated. This implies that the estimation of $r=0.2$ by BICEP-2 is not completely due to cosmological origin but probably due to dust foregrounds. Hence the value of $r$ is very likely to be lower than $0.2$. 
Results from the joint Planck+BICEP-2 analysis is required in anticipation of the value of $r$ accurately.

Due to these discrepancies, we consider two different cases for our analysis. First, we use the recent measurements from WP+Planck+BICEP-2. 
Secondly, we use only WP+Planck with a prior on $r_{0.002} \in [0, 0.1]$ and $r_{0.002} \in [0,0.01]$. This case explains the effect of lower value of $r$ on the PPL model. 

\subsection{Estimation with WP+ Planck + BICEP-2 likelihood}\label{secbicep}
For this analysis, we use WP+Planck+BICEP-2 likelihood, 
where we add up the results from commander\_v4.1\_lm49.clik,
lowlike\_v222.clik and CAMspec\_v6.2TN\_2013\_02\_26.clik likelihood
for WP and Planck \cite{wmap_like,Planck_like} and BICEP likelihood \cite{bicep_like} to perform parameter estimation using the cosmological parameter estimation code SCoPE \cite{scope}. We choose the $7$ parameters model as ($\Omega_{b}h^{2}$, $\Omega_{m}h^{2}$, $h$, $\tau, \nu_{st}$, $A_{s}$, $r$) for PPL model. 
\begin{figure}[h]
\centering
\includegraphics[width=6.0in,keepaspectratio=true]{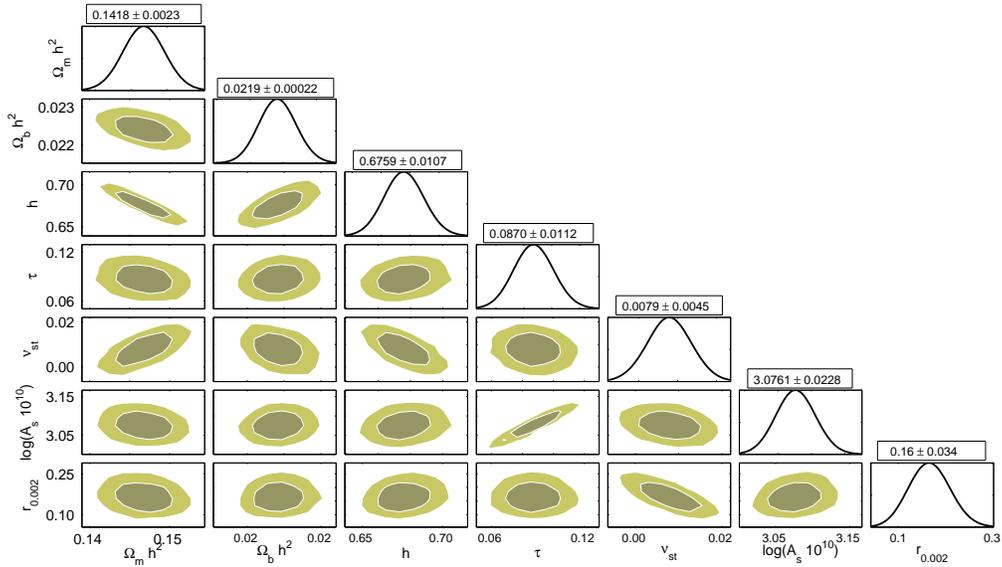}\label{nolen}
\caption{\label{fig:SP_2D}The two dimensional likelihood contours for the
$7$ parameters with $r_{0.002}$ for  WP+Planck+BICEP-2
likelihood are plotted in the lower triangle. The mean and standard deviation are mentioned for each parameters above the one dimensional marginalized probability distribution plot. 
These values are consistent with the results obtained by Planck \cite{Planck_param}.}\label{fig1}
\end{figure}

\begin{figure}[h]
\subfigure[]
{
\includegraphics[trim=0.30cm 0.12cm 0.25cm 0.15cm, clip=true, width=0.3\columnwidth]{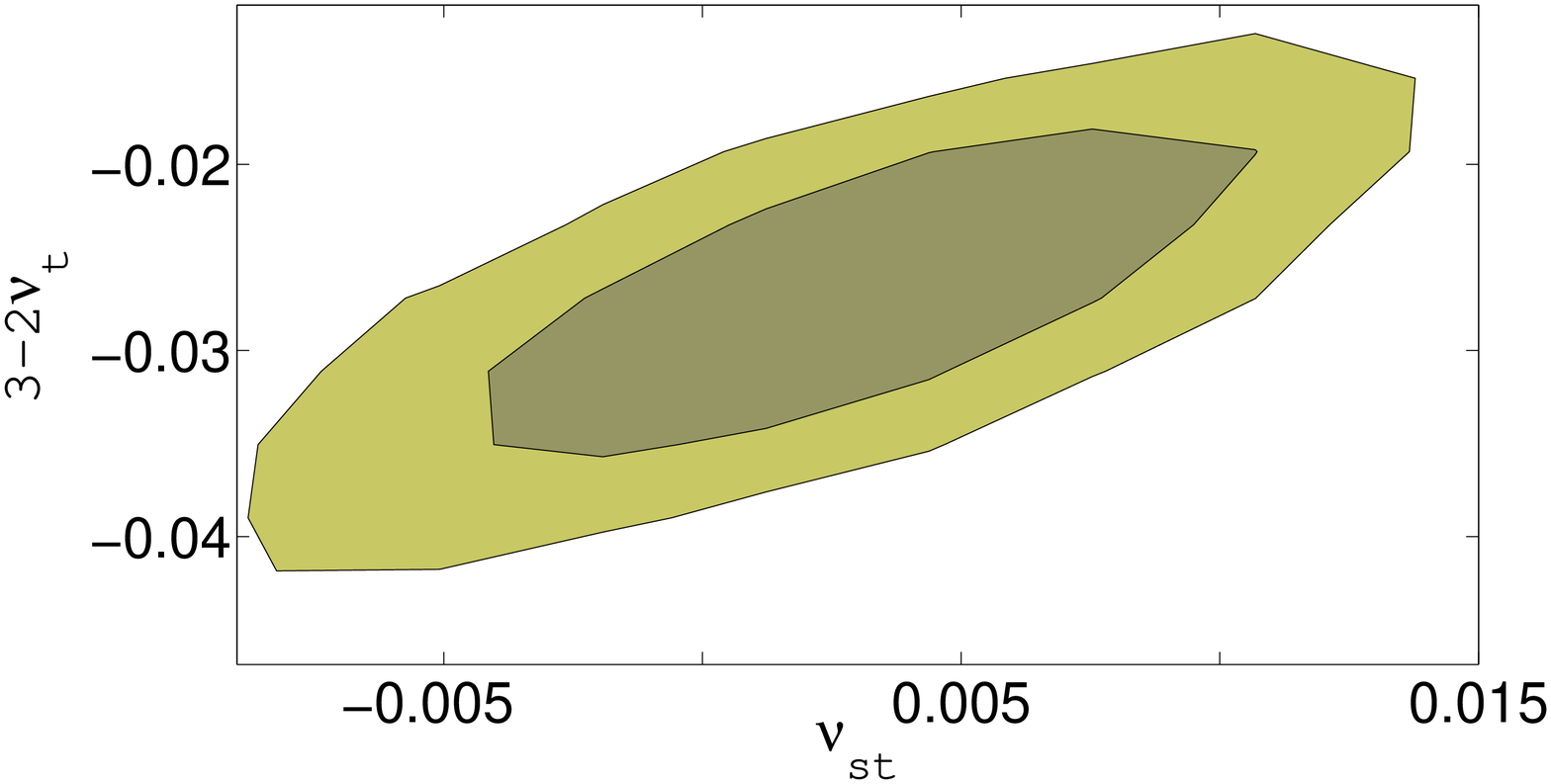}\label{ntpl}
}
\subfigure[]{
\includegraphics[trim=0.10cm 0.10cm 0.05cm 0.05cm, clip=true, width=0.3\columnwidth]{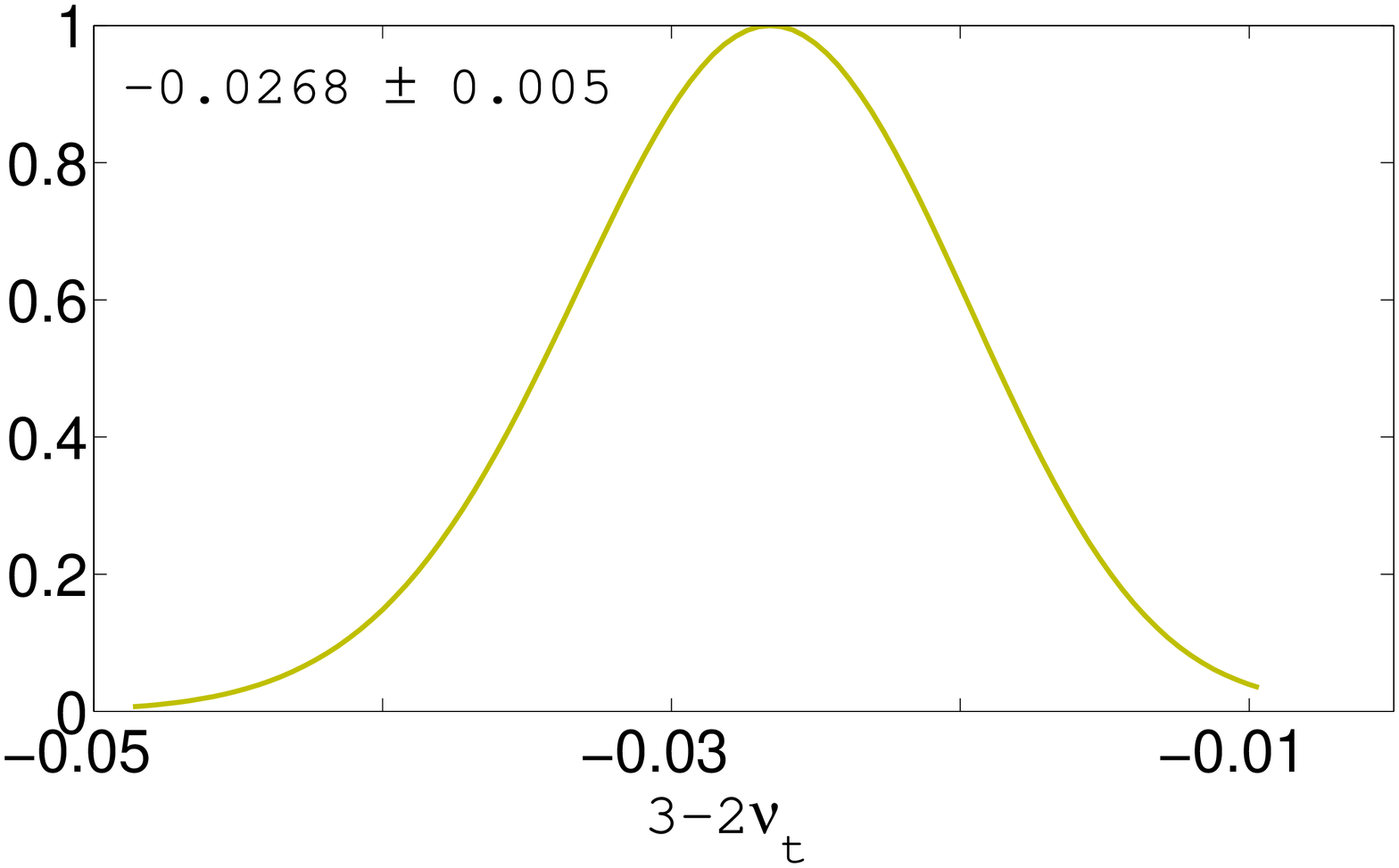}\label{nt1d}
}
\centering
\subfigure[]{
\includegraphics[trim=0.10cm 0.10cm 0.05cm 0.05cm, clip=true, width=0.3\columnwidth]{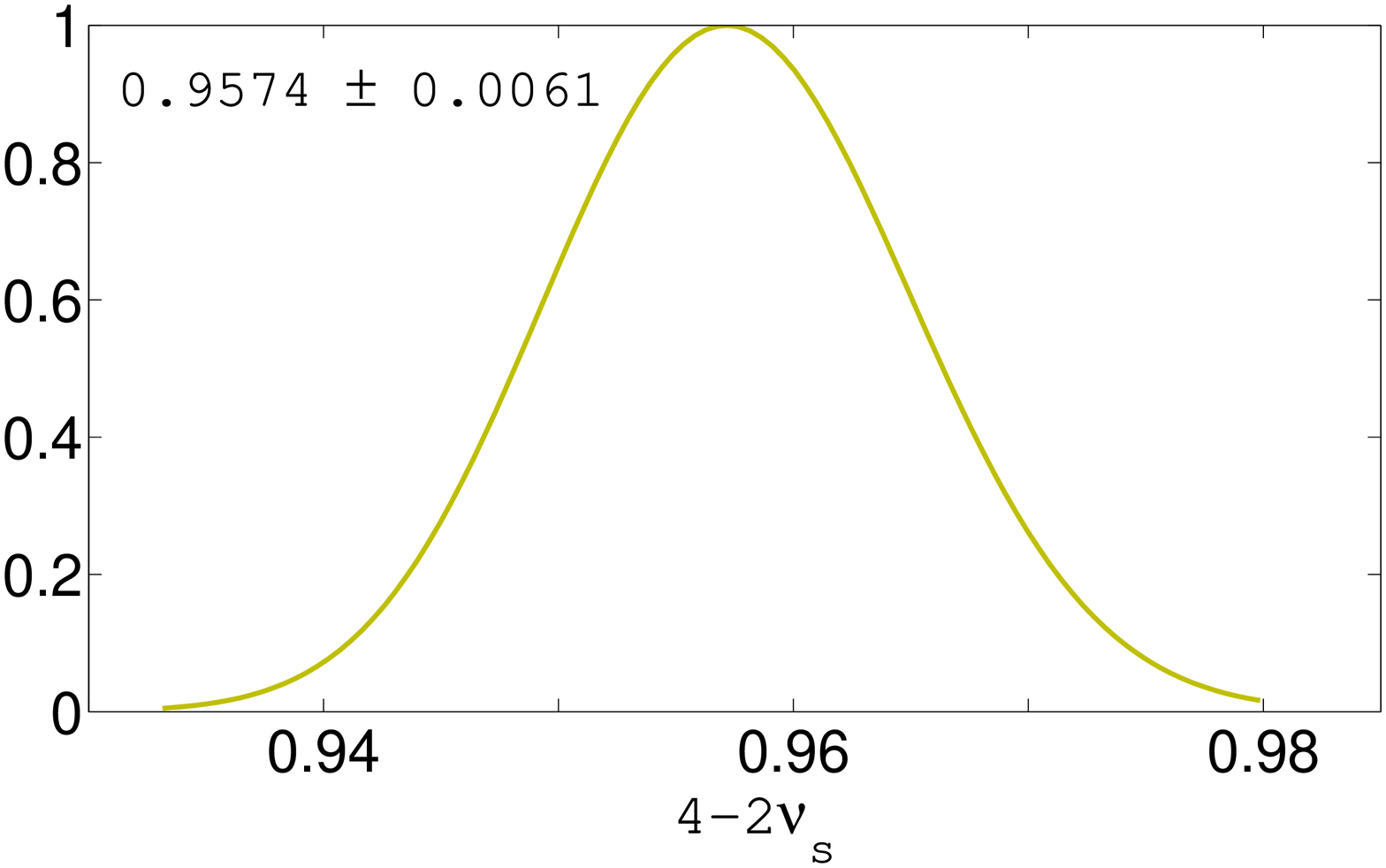}\label{ns1d}
}
\caption{\label{fig:SP_2D}(a): The two dimensional likelihood contours for the
$n_t=3- 2\nu_t$ and $\nu_{st}$ using WP+Planck+BICEP-2 likelihood. This shows a positive correlation between these two parameters and $\nu_{st}$ is consistent with zero within $95\,\%$ confidence limit. (b) One dimensional marginalized probability distribution for $n_t$ is plotted with the mean and the standard deviation. The data indicates non-zero  $n_t$ for PPL model with $5.36\sigma$.
 (c): One dimensional marginalized probability distribution for the derived parameter $n_s=4- 2\nu_s$ where $\nu_s=\nu_t +\nu_{st}$ is plotted with the mean and standard deviation to compare with the measurement of scalar spectral index $n_s$ obtained by Planck. This value is consistent within $1\sigma$ with the Planck measurement of $n_s = 0.9603 \pm 0.0073$ \cite{Planck_param}.}\label{fig2}
\end{figure}

Fig. \ref{fig1} gives the contour plots for the $7$ parameters estimated with PPL model. These parameters are completely consistent with the $\Lambda$CDM cosmological model supported by Planck \cite{Planck_param}. The mean for the inflationary parameter $\nu_{st} \equiv \nu_s -\nu_t = 0.0079$, which also constraints the effective mass $\frac{m_{\rm eff}^2}{\rm H^2} = -0.0237$. This measurement indicates that the departure from pure PL inflationary model is not significant ($1.76\sigma$). However, the negative value of $\frac{m_{\rm eff}^2}{\rm H^2}$ (or $\frac{d^2\ln \rm H}{d\phi^2}$) is more plausible with the data and indicates departure from PL inflation. For the other inflationary parameter $\nu_t$, we plot the two dimensional contour  between $n_t = 3-2\nu_t$ and $\nu_{st}$ given in Fig. \ref{ntpl} that also constraints $(\frac{d\ln \rm H}{d\phi})^2$ and $\frac{m_{\rm eff}^2}{\rm H^2}$ respectively. This plot indicates that the current data set (WP+Planck+BICEP-2) allows only a restricted feasible range of the inflationary  parameters to vary. In  Fig. \ref{nt1d} we plot the one dimensional distribution for $n_t$ with a mean of $-0.0268$. This show that for PPL model of inflation spectral index for tensor is non-zero at $5.36\sigma$. This result is expected to improve further from the polarization data of Planck.
It is essential to compare the scalar spectral index $n_s=4-2\nu_s$ obtained from PPL  model with the value obtained by Planck. As we have used the inflationary parameters $r$ and $\nu_{st}$ as free parameters, $\nu_s$ is obtained using 
$\nu_s = \nu_t +\nu_{st}$. We plot the one dimensional distribution for $n_s$ in Fig. \ref{ns1d} for the PPL model. The measurement is consistent with the result obtained by Planck ($n_s= 0.9603 \pm 0.0073$) \cite{Planck_param} within $1\sigma$. From this estimation we can conclude that the PPL model is consistent with current data set of WP, Planck and BICEP-2.  

\subsection{Estimation with WP+Planck likelihood}\label{secnobicep}
\begin{figure}[h]
\centering
\subfigure[]{
\includegraphics[width=6.0in,keepaspectratio=true]{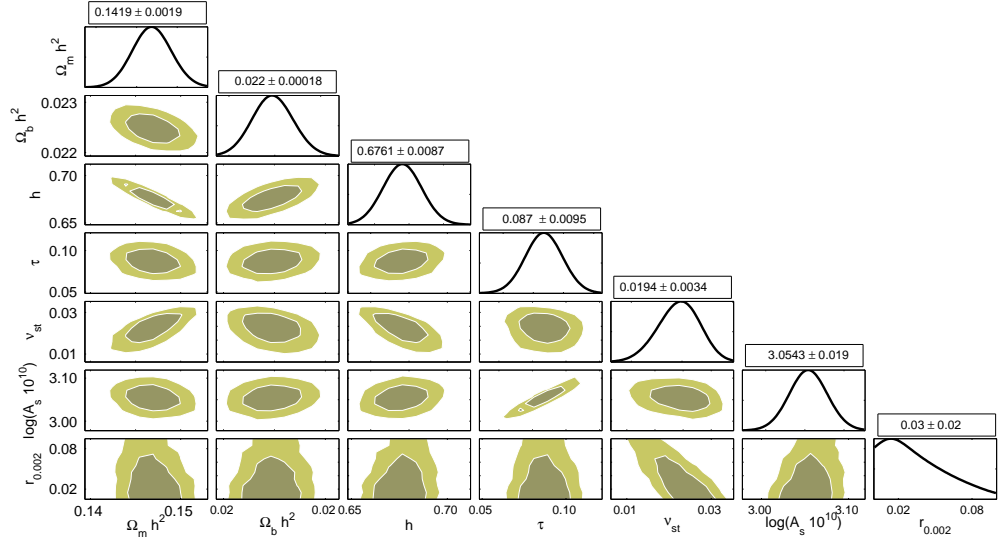}\label{nobicep1}
}
\subfigure[]{
\includegraphics[width=6.0in,keepaspectratio=true]{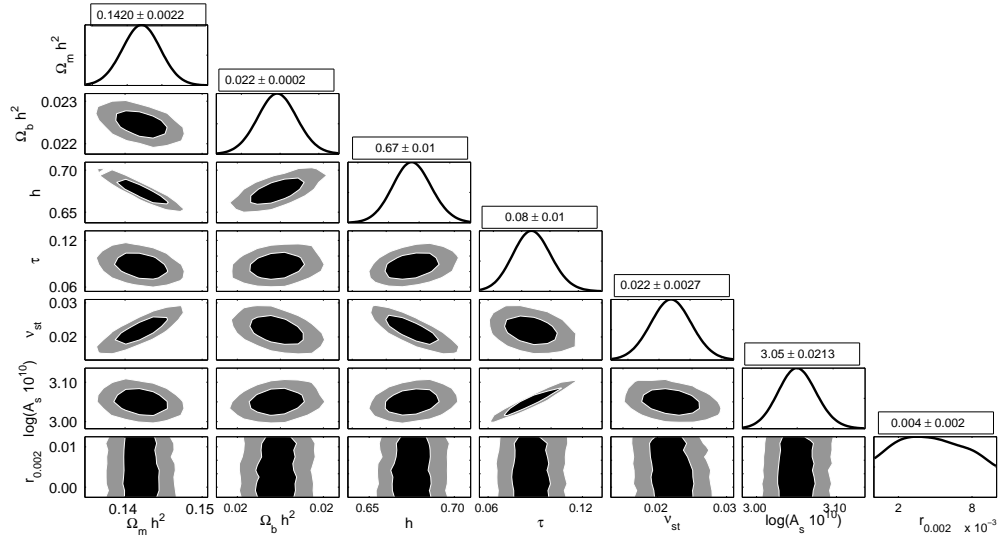}\label{nobicep2}
}
\caption{\label{fig:SP_2D}The two dimensional likelihood contours for the
$7$ parameters for  WP+Planck
likelihood are plotted in the lower triangle with prior on $r_{0.002}$ as (a) $[0, 0.1]$ and (b) $[0, 0.01]$. The mean and standard deviation are mentioned for each parameters above the one dimensional marginalized probability distribution plot.} \label{fig3} 
\end{figure}

\begin{figure}[h]
\subfigure[]
{
\includegraphics[trim=0.30cm 0.12cm 0.25cm 0.15cm, clip=true, width=0.3\columnwidth]{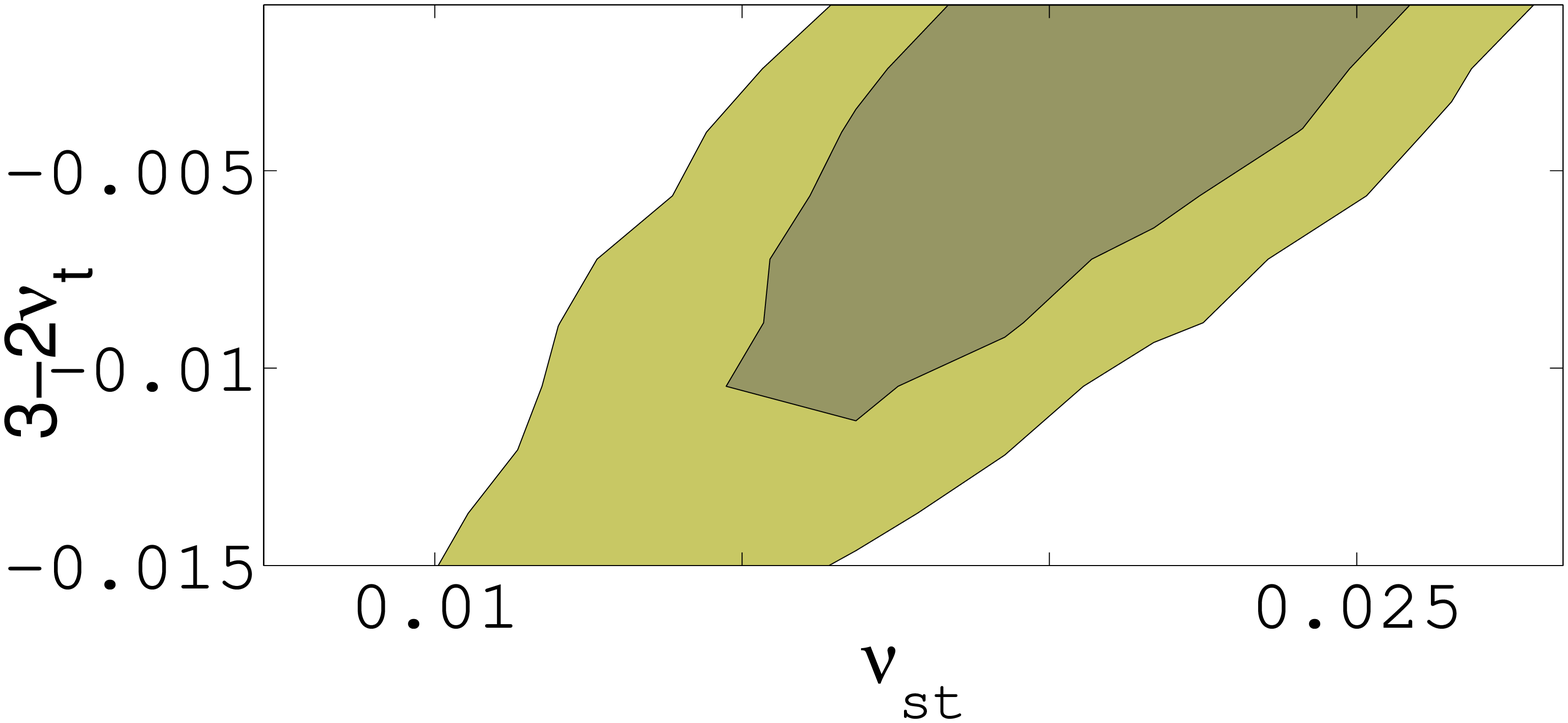}\label{ntpl1}
}
\subfigure[]{
\includegraphics[trim=0.10cm 0.10cm 0.05cm 0.05cm, clip=true, width=0.3\columnwidth]{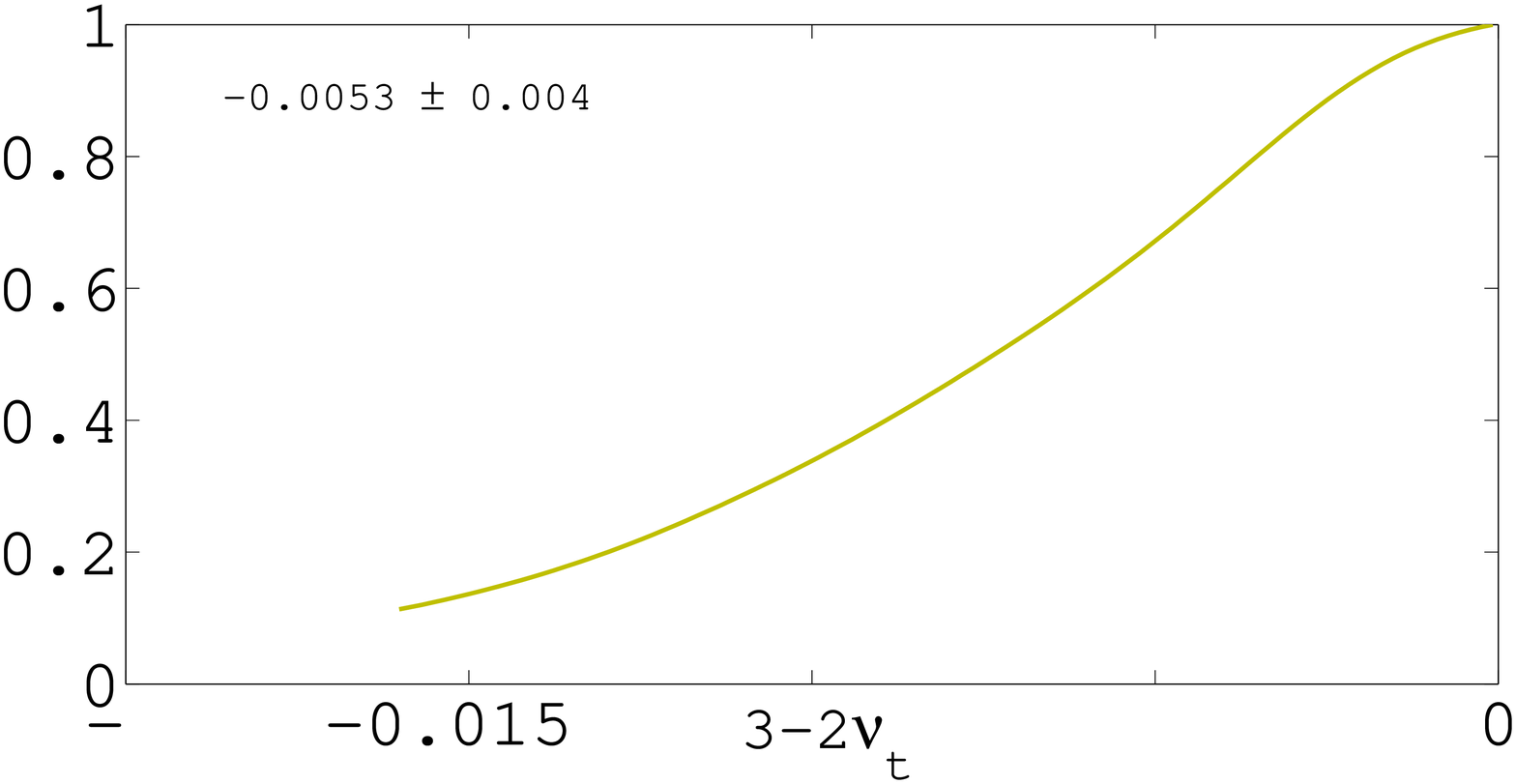}\label{nt1d1}
}
\centering
\subfigure[]{
\includegraphics[trim=0.10cm 0.10cm 0.05cm 0.05cm, clip=true, width=0.3\columnwidth]{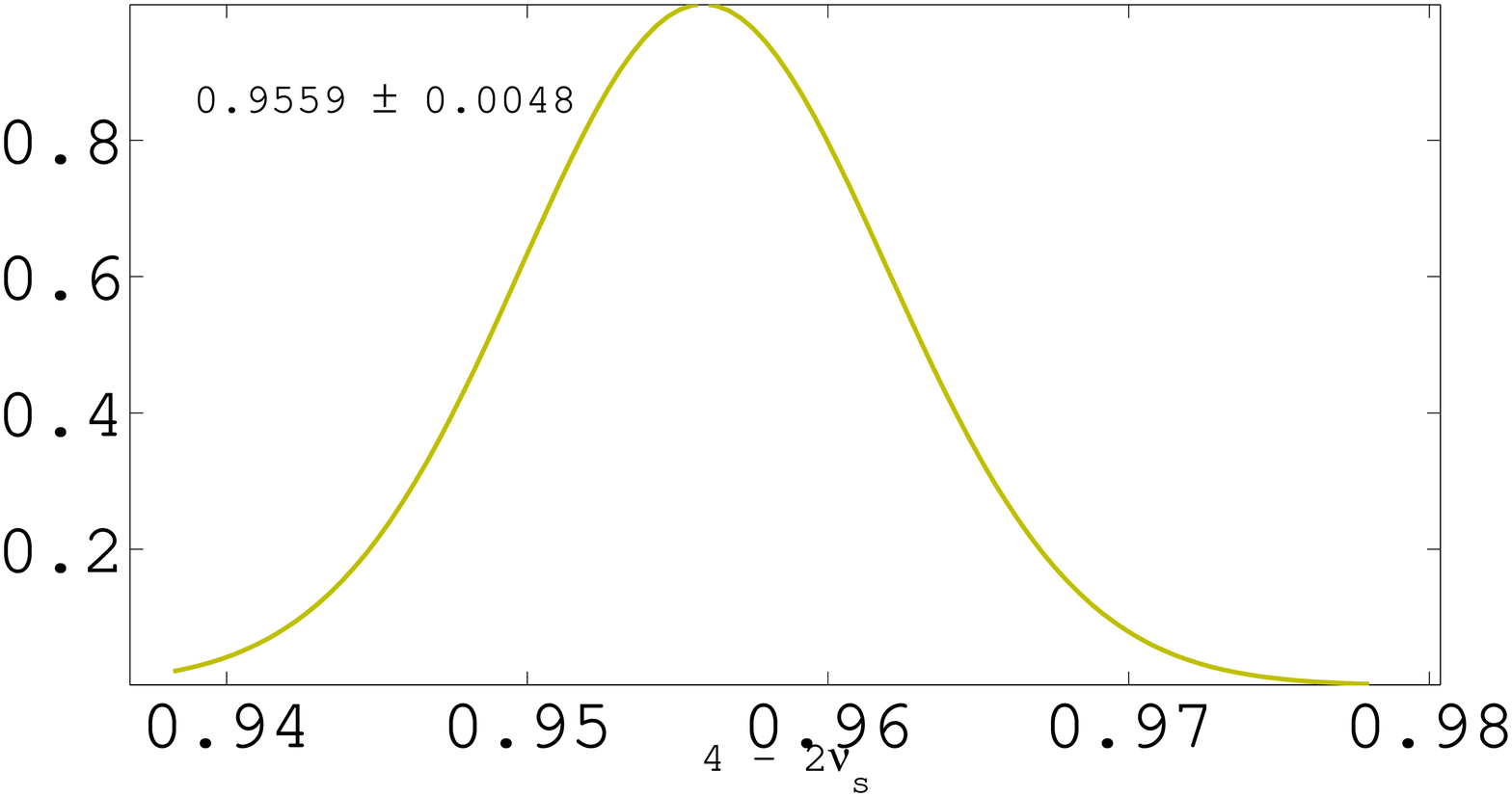}\label{ns1d1}
}
\subfigure[]
{

\includegraphics[trim=0.30cm 0.12cm 0.25cm 0.15cm, clip=true, width=0.3\columnwidth]{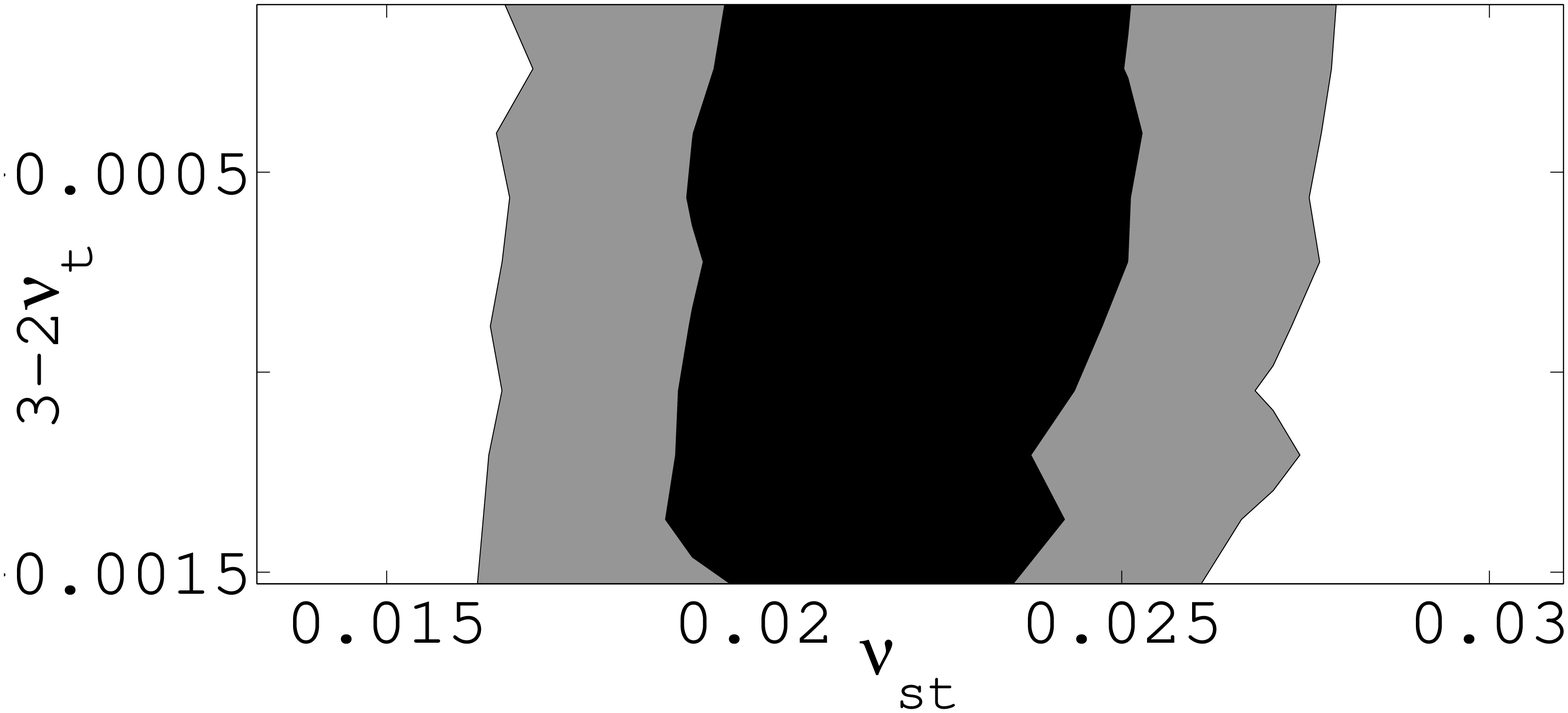}\label{ntpl2}
}
\subfigure[]{
\includegraphics[trim=0.10cm 0.10cm 0.05cm 0.05cm, clip=true, width=0.3\columnwidth]{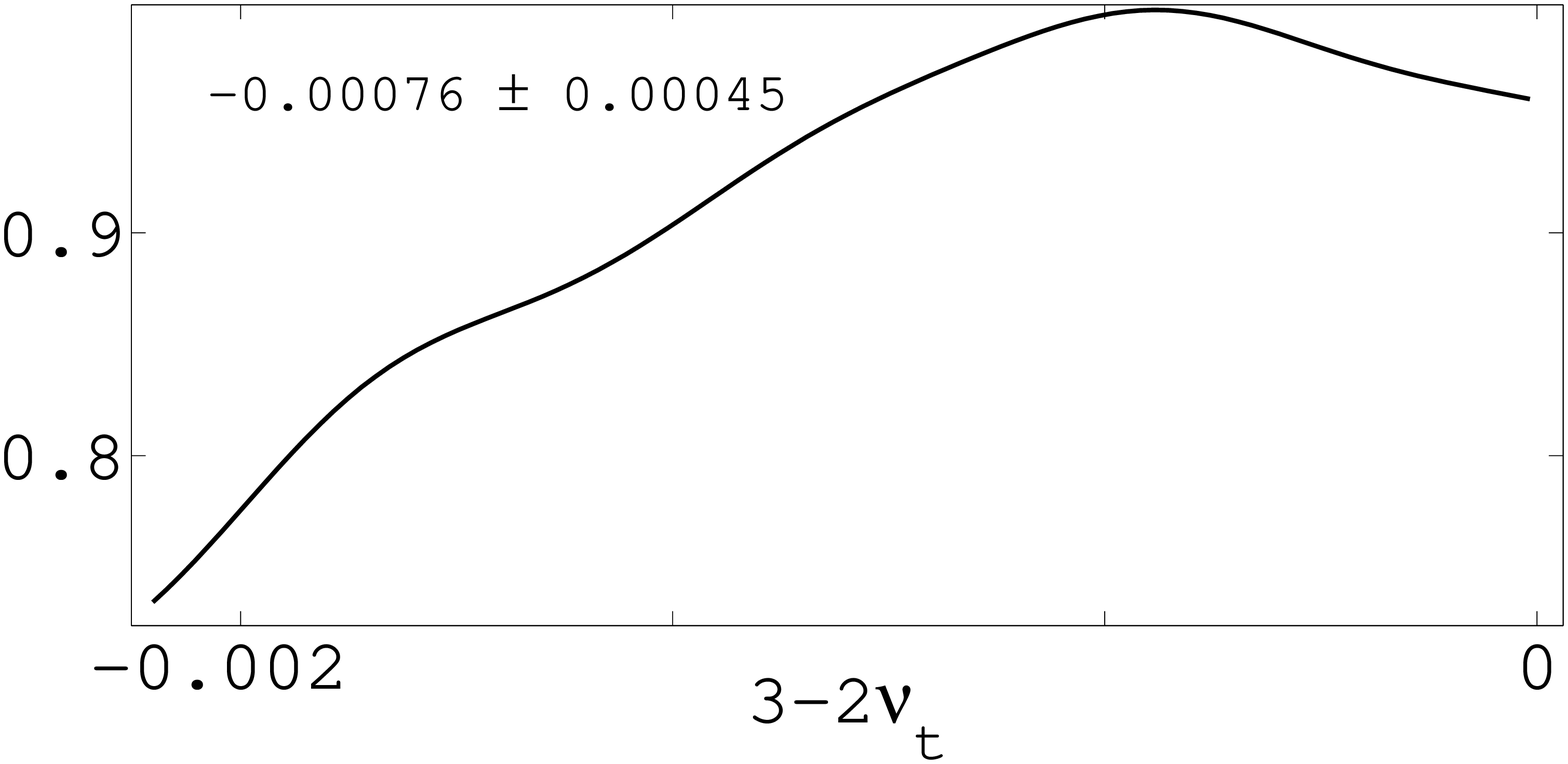}\label{nt1d2}
}
\centering
\subfigure[]{
\includegraphics[trim=0.10cm 0.10cm 0.05cm 0.05cm, clip=true, width=0.3\columnwidth]{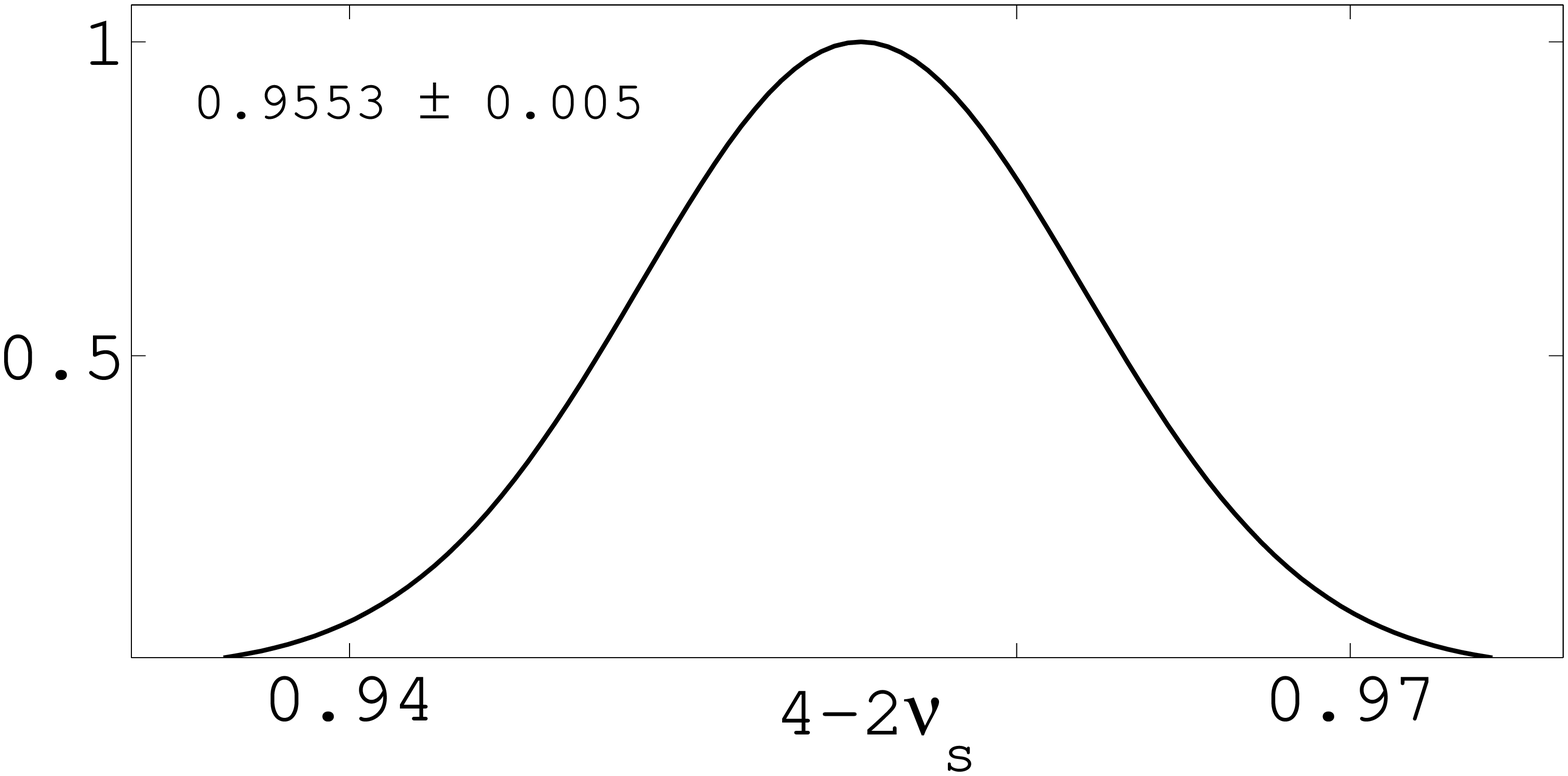}\label{ns1d2}
}
\caption{\label{fig:SP_2D}:  The two dimensional likelihood contours for the
$n_t=3- 2\nu_t$ and $\nu_{st}$ using WP+Planck likelihood with prior on (a) $r_{0.002} \in [0,0.1]$ and (d) $r_{0.002} \in [0,0.01]$. This shows a positive correlation between these two parameters and $n_{t}$ is consistent with zero within $95\,\%$ confidence limit. One dimensional marginalized probability distribution for $n_t$ is plotted with the mean and the standard deviation. The data indicates (b) $n_t= -0.005 \pm 0.004$ for $r_{0.002} \in [0,0.1]$ and (e) $n_t= -0.00076 \pm 0.0004$ for $r_{0.002} \in [0,0.01]$.
One dimensional marginalized probability distribution for the derived parameter $n_s=4- 2\nu_s$ where $\nu_s=\nu_t +\nu_{st}$ is plotted with the mean and standard deviation to compare with the measurement of scalar spectral index $n_s$ obtained by Planck. This value is consistent within $1\sigma$ with the Planck measurement of $n_s = 0.9603 \pm 0.0073$ \cite{Planck_param} for both (c) $r_{0.002} \in [0,0.1]$ and (f) $r_{0.002} \in [0,0.01]$.}\label{fig4}
\end{figure}

In this section, we use the likelihood only from WP+Planck and study two cases with the priors on $r_{0.002}$ as $[0, 0.1]$ and $[0, 0.01]$. We choose the same $7$ parameters model ($\Omega_{b}h^{2}$, $\Omega_{m}h^{2}$, $h$, $\tau, \nu_{st}$, $A_{s}$, $r$) to study the inflationary parameters for PPL model. In Fig. \ref{nobicep1} and \ref{nobicep2} we plot the contour plots for the parameters with prior on $r_{0.002} \in [0, 0.1]$ and $r_{0.002} \in [0, 0.01]$ respectively. All parameters except $\nu_{st}$ and $r_{0.002}$ do not change significantly in comparison to the previous analysis where we considered BICEP-2 likelihood. Due to lower value of $r$, the value of $n_t$ is high as shown in Fig. \ref{nt1d1} and Fig. \ref{nt1d2}. But the value of $n_s$ plotted in Fig. \ref{ns1d1} and Fig. \ref{ns1d2}, is nearly same for both the cases.  Hence the value of $\nu_{st}$ is larger than the value obtained in the previous section (Sec. \ref{secbicep}). For  $r_{0.002} \in [0, 0.1]$ and $r_{0.002} \in [0, 0.01]$, our analysis shows a non-zero measurement of  $\nu_{st}$ with $5.7\,\sigma$ and $8.1\,\sigma$ respectively. This also implies the value of $\frac{m^2_{\rm eff}}{\rm H^2}$ as $-0.006 \pm 0.001$ and $0.007 \pm 0.0009$ respectively. 

This part of the analysis shows that a lower value of $r_{0.002}$ with $n_s =0.9559$ can be easily incorporated in PPL model of inflation by a non-zero value of $\frac{m^2_{\rm eff}}{\rm H^2}$. In PL model, the value of $\frac{m^2_{\rm eff}}{\rm H^2}=0$, as a result of which  lower value of $r$ cannot be incorporated. This implies that PPL is a robust parametrisation  which is capable of accommodating the current uncertainties in $n_s$ and $r$.  

\section{Discussions and Conclusions}\label{conc}
Power Law (PL) inflationary model considers constant spectral index for both scalar and tensor perturbations as shown in Eq. \ref{nsnt}. This results into only non-zero $(\frac{d\ln \rm H}{d\phi})^2$ and all other higher order terms as zero. 
Perturbed Power Law (PPL) model of inflation introduced by Souradeep et al. \cite{ts},  is a soft departure from PL model by considering the higher order derivatives of Hubble parameter ($\frac{d^2\ln \rm H}{d\phi^2}$) during inflation. This term lead to varying spectral index for scalar perturbations but the constant spectral index for tensor perturbations and hence leads to observable features on CMB power spectra $C_l^{XX}$ for $X \in {T, \,E, \,B}$. In PPL model the feature of $n_s, n_t$ and $n_{run}$ can be incorporated by any two parameters from ${\nu_s, \nu_t, \nu_{st}}$. With the precision measurement of CMB  temperature and polarization power spectra, we can estimate these inflationary parameters which are important tools for understanding the inflationary era. 

We estimate the parameters for the PPL model for two different cases in our analysis, $(i)$ using BICEP-2 likelihood and $(ii)$ without considering the measurements form BICEP-2. Since, Planck have indicated a possible contamination in the polarization field from dust foregrounds in the BICEP-2 results and which can lead to a value of $r$ smaller than $0.2$. To understand the viability of the PPL model for small $r$ values, we have chosen two different priors on $r$ ($[0, 0.1]$ and $[0, 0.01]$)

Using the recent measurements of CMB temperature and polarization from WMAP, Planck and BICEP-2, we estimate the $7$ parameters cosmological model ($\Omega_{b}h^{2}$, $\Omega_{m}h^{2}$, $h$, $\tau, \nu_{st}$, $A_{s}$, $r$) for PPL inflation. In Fig. \ref{fig1}, we plot the contour and one dimensional plot for $7$ parameters and it shows that all the standard parameters are consistent with the PPL model. The inflationary parameter $\nu_{st}= \nu_s-\nu_t$ peaks at $0.0079$ and is consistent with zero at $1.76\sigma$. As shown in Eq. \ref{nsme}, this translates to effective mass $\frac{m^2_{\rm eff}}{\rm H^2}= -0.0237$. Though this value is consistent with zero within $2\sigma$, negative value of $\frac{m^2_{\rm eff}}{\rm H^2}$ seems more plausible with the data. The two dimensional contour for the inflationary parameters $\nu_t$ and $\nu_{st}$ is plotted in Fig. \ref{ntpl}. From the data we get non-zero value of $n_t= 3-2\nu_t = 0.0268$ with a $5.36\sigma$ significance shown in Fig. \ref{nt1d}. This estimation can be improved further with the future measurement of polarization by Planck and other experiments. We also estimate $\nu_s$ which is related to other inflationary parameters by $\nu_s = \nu_t +\nu_{st}$. In Fig. \ref{ns1d}, we show that $n_s=4-2\nu_s$ peaks at $0.9574$ and is consistent within $1\sigma$ with $n_s= 0.9603 \pm 0.0073$ measured by Planck \cite{Planck_param}.

The analysis with WP + Planck likelihood plotted in Fig. \ref{fig3} shows that the parameters are consistent with $\Lambda$CDM model. Only the value of $\nu_{st}$ and $r$ change significantly on considering WP+Planck likelihood in comparison with the WP+Planck+BICEP-2 likelihood. Our analysis shows that $\nu_{st}$ which is related to $\frac{d^2\ln \rm H}{d\phi^2}$ would have a non-zero detection with $5.7\,\sigma$ and $8.1\,\sigma$ respectively for two different prior cases $[0, 0.1]$ and $[0, 0.01]$ as shown in  Fig. \ref{fig3}.  This is due to the fact that, for a well determined $n_s=0.955 \pm 0.005$ (Fig. \ref{ns1d1} and Fig. \ref{ns1d2}), a lower value $r$ considered here, leads to a  larger discrepancy between $n_t$ and $n_s-1$.  
Using Eq. \ref{nsme}, we calculate the effective mass of inflationary field ($\frac{m_{eff}^2}{H^2}$) as, $-0.006 \pm 0.001$ and $-0.007 \pm 0.0009$ for $r_{0.002} \in [0, 0.1]$ and $r_{0.002} \in [0, 0.01]$ respectively. 


From the above discussions we can conclude that PPL model of inflation, which is a mild departure from PL can explain the lower value of $r<0.1$ for a $n_s$ of  $0.955$ by considering a non-zero value of $\frac{m_{eff}^2}{H^2}$.  With a better constraint on $r$ from the polarization measurements by Planck and other future missions, we can estimate the values of $m_{\rm eff}$ and $\nu_t$ more precisely, which will help in understanding the exact nature of the inflation.

\pagebreak[4]

\acknowledgments{
We acknowledge the use of HPC facility at IUCAA for the required computation.
S. M. and S. D. acknowledge Council for Science and Industrial Researchers
(CSIR), India, for the financial support as Senior Research Fellows. MJ acknowledges the Associateship of IUCAA.}

\end{document}